\documentclass[fleqn,usenatbib]{mnras}

\pdfoutput=1

\usepackage{newtxtext,newtxmath}

\usepackage[T1]{fontenc}

\DeclareRobustCommand{\VAN}[3]{#2}
\let\VANthebibliography\thebibliography
\def\thebibliography{\DeclareRobustCommand{\VAN}[3]{##3}\VANthebibliography}

\usepackage{graphicx}	
\usepackage{amsmath}	
\usepackage{bm}         

\title[BH Spin and Lense-Thirring Precession]{Constraining Black Hole Spin based on the Absence of Lense-Thirring Precession of Megamaser Clumps}

\author[Giner \& Loeb]{
Santiago Giner $^{1}$\thanks{E-mail: santiagoginer@college.harvard.edu}
and Abraham Loeb.$^{1}$\thanks{E-mail: aloeb@cfa.harvard.edu}
\\
$^{1}$Department of Astronomy, Harvard University, 60 Garden Street, Cambridge, MA 02138, USA.
}

\date{}

\pubyear{2021}

\begin{document}
\label{firstpage}
\pagerange{\pageref{firstpage}--\pageref{lastpage}}
\maketitle

\begin{abstract}
We make use of the absence of Lense-Thirring precession of the innermost megamaser clumps around black holes in order to place upper limits on the spin amplitude of seven black holes at the center of megamaser galaxies:  NGC 4258, NGC 2273, UGC 3789, NGC 3393, NGC 5495, NGC 5765b, and UGC 6093.  The constraints are  derived  by  requiring  that  the  associated precession  time-scale  be  longer  than  the  age  of  the  megamaser  clumps, in order to preserve the single-plane geometry of the megamasers. 
\end{abstract}

\begin{keywords}
black hole physics -- masers -- relativistic processes
\end{keywords}



\section{Introduction}
\label{sec:1}

General Relativity admits, in the approximation of weak gravitational fields and objects moving at non-relativistic speeds, an analogy with the classical theory of electromagnetism \citep{schafer_gravitomagnetic_2004, ruggiero_gravitomagnetic_2002}. In electromagnetism, the magnetic field arises as a direct consequence of combining Coulomb's Law with special relativity, by demanding that electric forces be consistent with the Lorentz transformations between different reference frames. In exactly the same way, "gravitomagnetic" fields arise from the movement of mass at the equivalent of mass "charge." One can then introduce gravitoelectric ($\mathbf{E}$) and gravitomagnetic ($\mathbf{H}$) fields, which (in the weak-field, non-relativistic approximation) satisfy an analogue of Maxwell's equations \citep{chashchina_elementary_2009, mashhoon_gravitoelectromagnetism_2008}. This inevitably leads to the introduction of a velocity-dependent, Lorentz force in the equations of motion.
\\

In the context of this theory, \cite{lense_uber_1918}, analyzed the gravitomagnetic field produced by a rotating solid sphere and found that the velocity-dependent force resulting from that rotation causes a precession of the angular momentum associated with the orbit of a test particle around the central sphere \citep[see also][]{mashhoon}. This effect is known as the Lense-Thirring precession. It has been utilized to place a constraint on the spin of SgrA*, based on the common orbital plane of stars near it \citep{fragione_upper_2020}. Here we use a similar approach for the megamaser clumps, which are observed to orbit on a common plane around black holes in extragalactic active galactic nuclei. This approach is justified if the clumps are associated with stars, as explained by \cite{milosavljevic_link_2004}.\\

In Section \ref{sec:2} we present the basic formulae describing the Lense-Thirring effect and in Section \ref{sec:3} we make use of it in order to constrain the spin of seven megamaser galaxies, describing our procedure and summarizing our results. In Section \ref{sec:4} we summarize our conclusions.

\section{Lense-Thirring Precession}
\label{sec:2}

The dimensionless spin parameter vector $\bm{\chi}$ of a rotating black hole (BH) of mass $M_{\rm{BH}}$ is defined from the Kerr metric as $\bm{\chi} = \bm{j}/(GM_{\rm{BH}}/c^2)$, where $\bm{j}$ is the normalized angular momentum per unit mass of the BH, namely, $\bm{j} = \bm{J}/M_{\rm{BH}}c$, and $\bm{J}$ is the spin angular momentum of the BH \citep{loeb_supermassive}, which can be expressed in terms of $\bm{\chi}$ as:

\begin{equation}
\label{eq:bh_angular_momentum}
    \mathbf{J} = {\bm{\chi}} \frac{G M_{\rm{BH}}^2}{c}.
\end{equation}

Spinning black holes also exhibit a quadrupole moment $\mathcal{Q}$, given by \citep{merritt_dynamics_2013}:

\begin{equation}
\label{eq:quadrupole}
  \mathcal{Q} = -\frac{J^2}{c^2 M_{\rm{BH}}}.  
\end{equation}

\begin{figure}
\begin{center}
   \includegraphics[scale=0.5]{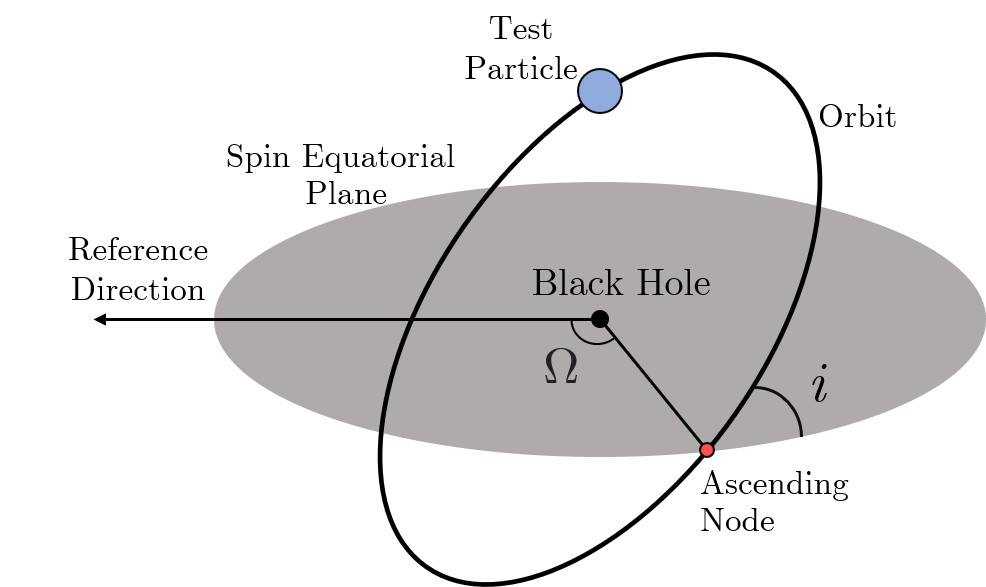}
\caption{A sketch of the orbital elements. The point at which the particle intersects the equatorial plane perpendicular to the BH spin, moving up, is the ascending node. The vertical tilt of the orbital plane with respect to the equatorial plane is the inclination $i$. The longitude of the ascending node $\Omega$ is the angle between an arbitrary reference direction and the ascending node.}
\label{fig:1}
\end{center}
\end{figure}

The Keplerian orbital parameters that we use in describing the orbit of test particles around the BH include the semi-major axis, $a$, eccentricity, $e$, the inclination, $i$, and the longitude of the ascending node, $\Omega$.
Figure \ref{fig:1} illustrates the geometry of the orbit. Both the spin angular momentum and the quadrupole moment induce a precession of the orbital plane of test particles around the BH. There is a secular precession of the longitude of the ascending node and the argument of periapsis, but the latter does not perturb the single-plane structure of megamasers, given that their orbits have negligible eccentricity \citep{kuo_megamaser_2011}. The associated Lense-Thirring precession does not affect the semi-major axis and only changes the eccentricity and the inclination in a periodic manner \citep{mashhoon, merritt_dynamics_2013, merritt_testing_2010}, leaving them unchanged when averaged over the orbital period. The change in the longitude of the ascending node due to the spin of the black hole, $\Delta\Omega_J$, where the subscript $J$ denotes the effect due to the BH spin, over one orbital period is \citep{merritt_dynamics_2013}

\begin{equation}
\label{eq:delta_Omega1}
    \Delta \Omega_J = \frac{4\pi \chi}{c^3}\left[\frac{GM_{\rm{BH}}}{a(1-e^2)}\right]^{3/2}.
\end{equation}
Similarly, the contribution from the BH quadrupole moment, $\Delta \Omega_{\mathcal{Q}}$, is

\begin{equation}
\label{eq:delta_Omega2}
    \Delta \Omega_{\mathcal{Q}} = \frac{3 \pi \chi^2}{c^4}\left[\frac{GM_{\rm{BH}}}{a(1-e^2)}\right]^{2} \cos{i}.
\end{equation}
Equations \eqref{eq:delta_Omega1}, \eqref{eq:delta_Omega2} hold for the case in which the BH spin axis is aligned with the $z$-axis of the coordinate system, which in fact does not imply a loss of generality \citep{merritt_dynamics_2013, merritt_testing_2010}; see, e.g., \cite{iorio_post-keplerian_2017} for an explicit derivation in the case of a BH spin axis misaligned relative to the line of sight.\\

The precession induced by the BH quadrupole moment is negligible with respect to the precession induced by the BH spin. Consequently, we take into account only the contribution from the BH spin to the precession of the longitude of the ascending node.\\

The orbits of maser clumps can be approximated as Keplerian \citep{kuo_megamaser_2011}, with an orbital period $P$,

\begin{equation}
\label{eq:period}
    P = 2\pi\left(\frac{a^3}{GM_{\rm{BH}}}\right)^{1/2}.
\end{equation}

We can therefore rewrite equation \eqref{eq:delta_Omega1} as

\begin{equation}
    \Delta \Omega_J = \frac{2 \chi G^2 M_{\rm{BH}}^2}{c^3 a^3(1-e^2)^{3/2}}P.
\end{equation}

Averaging over one period, we obtain the orbit-averaged time derivative of the longitude of the ascending node \citep{merritt_dynamics_2013},

\begin{equation}
\label{eq:d_Omega}
    \bigg \langle\frac{d\Omega_J}{dt} \bigg\rangle = \frac{2 \chi \: G^2 M_{\rm{BH}}^2}{c^3 a^3(1-e^2)^{3/2}}.
\end{equation}

Assuming that the eccentricity is negligible \citep{kuo_megamaser_2011}, equation \eqref{eq:d_Omega} becomes,

\begin{equation}
\label{eq:orbit-avg-Omega}
    \bigg \langle\frac{d\Omega_J}{dt} \bigg\rangle = \frac{2 \chi \: G^2 M_{\rm{BH}}^2}{c^3 a^3}.
\end{equation}

From equation \eqref{eq:orbit-avg-Omega}, we can obtain the Lense-Thirring precession period:

\begin{equation}
\label{eq:t_lt}
    T_{\rm{LT}} = \frac{c^3 a^3}{2 G^2 M^2 |\chi|},
\end{equation}
where the absolute value was added since the spin parameter $\chi$ could be negative for retrograde orbits.

\section{Data and Results}
\label{sec:3}

The Lense-Thirring precession time-scale is proportional to the cube of the semi-major axis of the orbit. Therefore, in the case of maser disks, test particles that are closer to the central BH will possess a shorter precession timescale, i.e. their orbit will be affected faster than clumps farther away. We can require that the Lense-Thirring precession time-scale given by equation \eqref{eq:t_lt} be greater than the age of the maser clumps $T$. The reason behind this constraint is that if the precession time-scale is smaller than the age of the clumps, then their orbits are going to precess in their lifetimes, causing the clumps on the innermost orbital radius to be misaligned with respect to those farther out. This would break the single-plane geometry. Thus,

\begin{equation}
\label{eq:condition}
    T_{\rm{LT}} > T \implies \frac{c^3 a^3}{2 G^2 M_{\rm{BH}}^2 |\chi|} > T.
\end{equation}

By solving inequality \eqref{eq:condition} we are able to obtain an upper limit on the spin amplitude:

\begin{equation}
\label{eq:constraint}
    |\chi_{\rm{max}}|< \frac{c^3 a^3}{2 G^2 M^2_{\rm{BH}} T}.
\end{equation}

\begin{table}
\centering
\caption{Upper limit on the spin amplitude, $|\chi_{\rm{max}}|$. Col. (1): Galaxy name. Col. (2): Mass of the black hole. Col. (3): Innermost orbital radius of maser clumps. Col. (4): Upper limit on spin parameter of the black hole. References for col. (2) and col. (3): Gao et al. (2016b).}

\label{table:1}
 \begin{tabular}{c c c c} 
 \hline
 \hline
  Name & BH mass ($10^7 M_\odot$) & $R_{\rm{in}} (\rm{pc})$ & $|\chi_{\rm{max}}|$ \\ [0.8ex]
 \hline
 NGC 4258 & $4.00 \pm \: 0.09$  & $0.11 \pm \: 0.004$ & 0.06 \\ 
 \hline
 NGC 2273 & $0.75 \pm \: 0.05$  & $0.03 \pm \: 0.01$ & 0.03 \\ 
 \hline
  UGC 3789 & $1.04 \pm \: 0.06$  & $0.08 \pm \: 0.02$ & 0.34 \\ 
 \hline
  NGC 3393 & $3.10 \pm \: 0.37$  & $0.17 \pm \: 0.02$ & 0.36 \\ 
 \hline
  NGC 5495 & $1.05 \pm \: 0.20$  & $0.10 \pm \: 0.05$ & 0.65 \\ 
 \hline
  NGC 5765b & $4.55 \pm \: 0.31$  & $0.30 \pm \: 0.06$ & 0.93 \\ 
 \hline
   UGC 6093 & $2.65 \pm \: 0.23$  & $0.12 \pm \: 0.07$ & 0.18 \\ 
 \hline
 \hline
 \\
\end{tabular}
\end{table}

Next, we use the tabulated data from \cite{gao_megamaser_2016}, see Table \ref{table:1}, regarding the innermost orbital radius of the observed maser clumps and the BH mass for seven megamaser galaxies where the constraints on the spin turn out to be compelling: NGC 4258, NGC 2273, UGC 3789, NGC 3393, NGC 5495, NGC 5765b, and UGC 6093.\\

\begin{figure}
\begin{center}
   \includegraphics[scale=0.65]{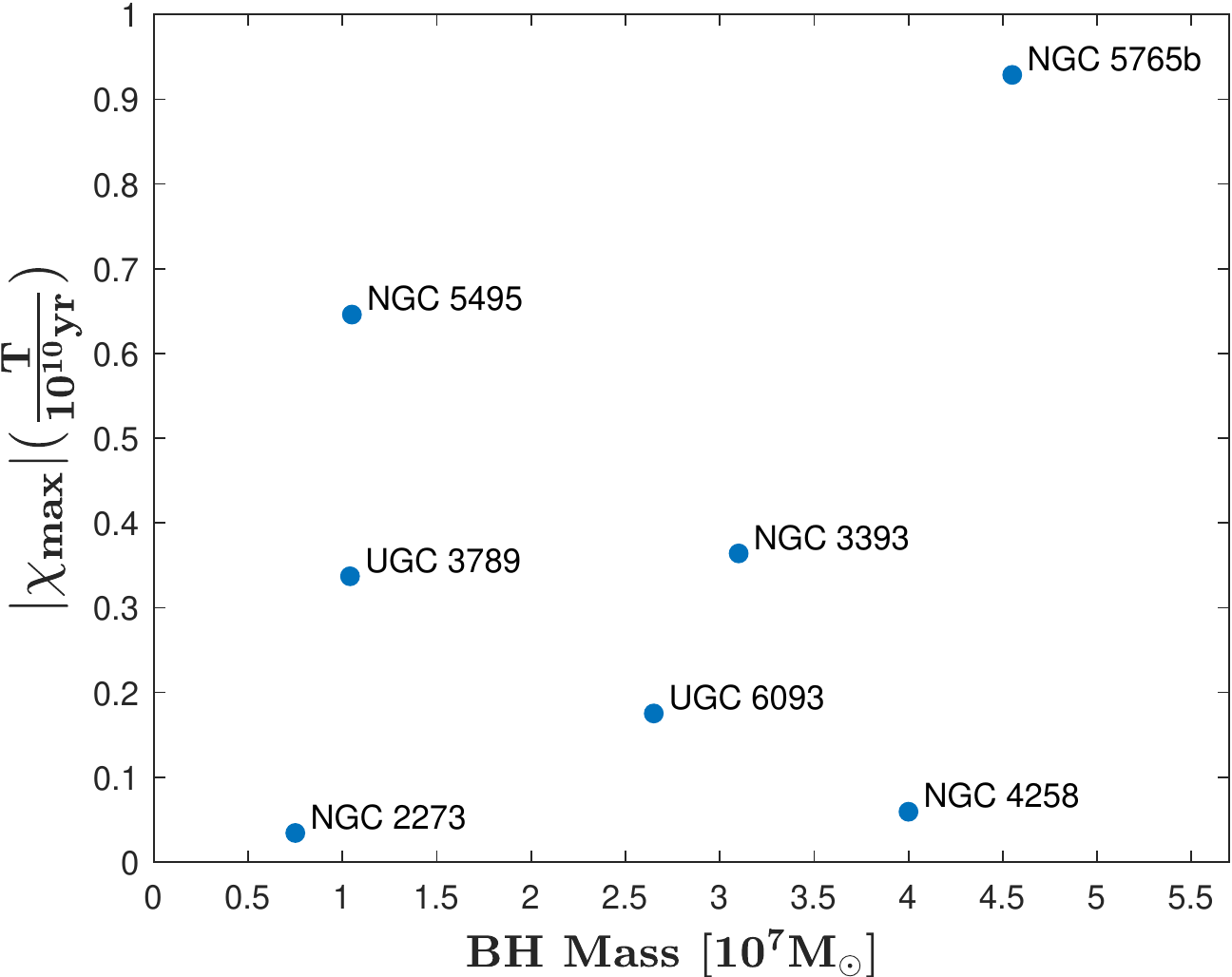}
\caption{Maximum BH spin amplitude, $|\chi_{\rm{max}}|$, for a maser clump's lifetime $T$ in units of $10^{10}$ yr versus BH mass. The limit on the spin amplitude is plotted along with the time $T$ because of the uncertainty in the age of the clumps.}
\label{fig:2}
\end{center}
\end{figure}

 The age $T$ in equation \eqref{eq:constraint} can go up to the age of the host galaxies, which, as stated in \cite{kuo_megamaser_2011}, is $\sim 10^{10}$ yr. The resulting spin parameter limit $|\chi_{\rm{max}}|$ is displayed for $T = 10^{10}$ yr in Table \ref{table:1}. Figure \ref{fig:2} shows the same limits $|\chi_{\rm{max}}| (T/10^{10}$ yr) as a function of the BH mass in the different galaxies. We have plotted the upper limit on the spin amplitude together with the time $T$ in order to emphasize the uncertainty regarding the age of the maser clumps. In that regard, better estimations of $T$ would provide for better limits on the spin amplitude (see Section~\ref{sec:4}). Notably, if the age estimation of the megamaser clumps is closer to $\sim 10^8$ yr, the constraints on the spin amplitude are no longer meaningful. Therefore, only for large values of $T$ can the spin be constrained in this way.

\section{Conclusion}
\label{sec:4}

We derived upper limits on the magnitude of the spin parameter of seven BHs at the center of megamaser galaxies. The limits were derived by requiring that the Lense-Thirring precession timescale associated with the longitude of the ascending node of the orbits around the BH not have sufficient time to induce a misalignment of the maser clumps at the innermost orbital radius. The tightest constraint was obtained for NGC 2273, with $|\chi_{\rm{max}}| \lesssim 0.03$ for $T \sim 10^{10}$ yr. The second tightest limit is for the spin amplitude of NGC 4258, $|\chi_{\rm{max}}|(T/10^{10}\rm{yr}) \lesssim 0.06$.\\

Our analysis was based on the assumption that megamaser clumps are much denser than their gaseous environment and therefore behave as test particles, which is justified if they are associated with stars, as suggested by \cite{milosavljevic_link_2004}. However, in order to explain the warping observed in the outer radius of the maser disk in NGC 4258 \citep{herrnstein_geometry_2005}, other studies have made use of the Bardeen-Petterson effect \citep{bardeen_lense-thirring_1975} in which a combination of Lense-Thirring precession and the viscosity of the gaseous disk lead to an alignment of the disk's angular momentum with the spin of the BH. These studies \citep[e.g.][]{franchini_lense-thirring_2015, caproni_bardeenpetterson_2006, caproni_is_2007} apply when the dynamics of the clumps is strongly affected by the surrounding gas.\\

Better estimates of the ages of the maser clumps might be obtainable from future spectroscopic observations of the stars in their vicinity with the James Webb Space Telescope or the next generation of ground-based large telescopes \citep{gullieuszik_probing_2014, martins_fundamental_2019}.

\section*{Acknowledgements}

We thank Giacomo Fragione, Dom Pesce, and Mark Reid for useful comments on the manuscript and interesting discussions.

\section*{Data Availability}

The data utilized for the numerical calculations in this paper can be found in the work of \cite{gao_megamaser_2016} and references therein. Extensive information on masers can be found in the series of papers from the Megamaser Cosmology Project \citep[e.g.][]{zhao_megamaser_2018, gao_megamaser_2015, kuo_megamaser_2011}.



\bibliographystyle{mnras}
\bibliography{main}



\bsp	
\label{lastpage}
\end{document}